\documentclass[twocolumn,aps,prb,showpacs]{revtex4}
\usepackage{amsmath}
\usepackage{amssymb}
\usepackage[tight,TABTOPCAP]{subfigure}
\usepackage{subfloat}
\usepackage{float}
\usepackage{graphicx}
\usepackage{dcolumn}
\usepackage{bm}
\usepackage{setspace}
\makeatletter

\begin{document}

\title{Signatures of Fermi surface reconstruction in Raman spectra of underdoped cuprates}

\author{J. P. F. LeBlanc$^{1,2}$}
\email{leblanc@physics.uoguelph.ca}
\author{J. P. Carbotte$^{3,4}$}
\author{E. J. Nicol$^{1,2}$}%
\affiliation{$^1$Department of Physics, University of Guelph,
Guelph, Ontario N1G 2W1 Canada} 
\affiliation{$^2$Guelph-Waterloo Physics Institute, University of Guelph, Guelph, Ontario N1G 2W1 Canada}
\affiliation{$^3$Department of Physics and Astronomy, McMaster
University, Hamilton, Ontario L8S 4L8 Canada}
\affiliation{$^4$The Canadian Institute for Advanced Research, Toronto, ON M5G 1Z8 Canada}
\date{\today}
\begin{abstract}
We have calculated the Raman B$_{1g}$ and B$_{2g}$ spectra as a function of temperature, as well as doping, for the underdoped cuprates, using a model based on the resonating valence-bond spin-liquid.  We discuss changes in intensity and peak position brought about by the presence of a pseudogap and the implied Fermi surface reconstruction, which are elements of this model.  Signatures of Fermi surface reconstruction are evident as a sharp rise in the doping dependence of the antinodal to nodal peak ratio which occurs below the quantum critical point.  The temperature dependence of the B$_{1g}$ polarization can be used to determine if the superconducting gap is limited to the Fermi pocket, as seen in angle resolved photoemission spectroscopy, or extends beyond.  We find that the slope of the linear low energy B$_{2g}$ spectrum maintains its usual d-wave form, but with an effective gap which reflects the gap amplitude projected on the Fermi pocket.  Our calculations capture the main qualitative features revealed in the extensive data set available on the HgBa$_2$CuO$_{4+\delta}$ (Hg-1201) cuprate.
\end{abstract}

\maketitle
\section{Introduction}
The underdoped region of the high T$_c$ cuprate phase diagram is known to contain a number of features which makes both the normal and superconducting state challenging to understand using microscopic models.  These features included: nanoscale inhomogeneities as seen in scanning tunneling microscopy (STM),\cite{pan:2001,mcelroy:2005} the observation of a pseudogap energy scale above T$_c$,\cite{loram:1998,renner:1998} as well as a transition to an antiferromagnetic Mott insulating state at low doping.\cite{lee:2006}

The discovery of a pseudogap feature in the normal state above the superconducting dome, for doping below some critical value (thought to be a quantum critical point), has recently resulted in considerable research activity.  Understanding the effect of this pseudogap energy scale on the transition temperatures, T$_c$, of the cuprates, as well as the origin of this energy scale itself, are important and interesting issues.
Examination of the cuprate phase diagram shows a clear reduction in T$_c$ as the pseudogap energy scale increases.  This could be an indication of the pseudogap arising from a competing phase, driving down the superconducting gap, and hence the T$_c$.  Indeed, a great deal of experimental evidence appears to suggest that the superconducting gap and pseudogap occupy unique regions of phase space (nodal superconductivity and antinodal pseudogap).\cite{yang:2008}

Since its appearance in 2006, the model of Yang, Zhang and Rice (YRZ)\cite{yrz:2006} has shown some success in understanding Raman and optical properties,\cite{valenzuela:2007,illes:2009} angle-resolved photoemission (ARPES)\cite{yrz:2009} and specific heat data \cite{leblanc:2009}.  The YRZ model provides an ansatz for the coherent part of the many-body Green's function for an resonating valence bond (RVB) spin liquid system.  In the absence of a pseudogap, the model maintains a large tight-binding Fermi surface that can undergo a transition to the superconducting state resulting in a superconducting gap which opens on this Fermi surface.  The inclusion of a pseudogap in the YRZ model is such that it opens about the antiferromagnetic Brillouin zone (AFBZ) boundary.  The net result is that a finite pseudogap in this model acts to deform the tight-binding Fermi surface to form Fermi hole pockets.  If one assumes the onset of the pseudogap to be a zero temperature quantum critical point (QCP) at a critical doping $x_c$, then we can define several regions around this critical doping.  First, for dopings above $x_c$, there is a superconductivity-dominated region with a large tight-binding Fermi surface.  Second, for dopings well below $x_c$, there is a pseudogap-dominated region with Fermi pockets which become smaller and shrink towards the nodal direction as the antiferromagnetic Mott insulator is approached. Finally, there is an intermediate region wherein the features of a pseudogap onset, as well as a deformation of Fermi surface both present themselves.  It is in this region that the energy scale of superconductivity and the pseudogap are comparable.

There are other theoretical methods for including a pseudogap phase through, for example, preformed pairs,\cite{emery:1995} or as a competing $d$-density wave order.\cite{chakravarty:2001}  As the YRZ theory has thus far resulted in good qualitative descriptions of a number of experimental properties, continued investigation of this model allows us to distinguish quintessential features of the cuprates which must be included in any more complex microscopic models that might follow.  Of central importance to this paper is the presence of two energy scales, nodal and antinodal, in the superconducting underdoped cuprates, as seen in Raman and angle-resolved photoemission spectroscopy (ARPES) experiments.\cite{kanigel:2006,guyard:2008,lee:2007,kondo:2009}  Within the model of YRZ, which includes two independent energy scales (superconducting and pseudogap), we calculate the minima in the angular band energies for comparison with ARPES and application to Raman spectra.  
Although the YRZ model has previously been applied to the calculation of Raman spectra at zero temperature in the work of Valenzuela \emph{et al.} \cite{valenzuela:2007}, the work presented here extends upon those doping dependent calculations to understand better the specific behaviours which are present in models of the pseudogap phase.
Here we will identify the impact on Raman B$_{1g}$ (antinodal) and B$_{2g}$ (nodal) spectra of two important behaviours: (1) the $k$-space separation of superconducting and pseudogap energies and (2) the reconstruction of the Fermi surface.   We will demonstrate how (1) and (2) play a role in the temperature and doping dependence of the Raman spectra.

We have organized this work to fit the following structure.  Section II will describe the theoretical framework and parameters involved in the YRZ model used for all calculations shown.  Section III will be a description of ARPES results in the context of the YRZ model with the motivation that this will be necessary for the understanding of the Raman results.  Section IV will contain our Raman calculations in the YRZ model, including finite temperature, while Section V will contain a summary of our main conclusions.
\section{Theory}

In the YRZ model,\cite{yrz:2006,yrz:2009} both the superconducting gap, $\Delta_{\rm sc}$, and
the pseudogap, $\Delta_{\rm pg}$, have a $d$-wave $k$-space dependence
described by: 
\begin{eqnarray}
\Delta_{\rm sc}=\frac{\Delta_{\rm sc}^{0}(x)}{2}(\cos k_xa -\cos k_ya)\label{eqn:scgap},\\
\Delta_{\rm pg}=\frac{\Delta_{\rm pg}^{0}(x)}{2}(\cos k_xa -\cos k_ya)\label{eqn:pggap},
\end{eqnarray}
where $a$ is the lattice constant.
For a doping $x$, the YRZ model employs, for the coherent piece, a propagator,
\begin{eqnarray}
G(\boldsymbol{k},\omega,x)=
\sum_{\alpha=\pm}  \frac{{g_t W^{\alpha}_{\boldsymbol{k}}}}{
{\omega-E^{\alpha}_{\boldsymbol{k}}-\Delta_{\rm sc}^{2}/(\omega+E^{\alpha}_{\boldsymbol{k}})}},
\label{eqn:sc}
\end{eqnarray}
which has been formulated from numerical studies of RVB-type models and will contain the essential physics of the superconducting state.  Entering this Green's function are the quantities:
\begin{align}
E_{\boldsymbol{k}}^ \pm  &= \frac{{\xi_{\boldsymbol{k}}  - \xi_{\boldsymbol{k}}^0 }}{2} \pm E_{\boldsymbol{k}}, \nonumber \\
E_{\boldsymbol{k}} &= \sqrt {\tilde{\xi}_{\boldsymbol{k}}^2  + \Delta
  _{\rm pg}^2 },\nonumber \\
\tilde{\xi}_{\boldsymbol{k}}  &= \frac{(\xi_{\boldsymbol{k}}  + \xi_{\boldsymbol{k}}^0 )}{2}, \nonumber \\
W_{\boldsymbol{k}}^ \pm  &= \frac{1}{2} \left(1\pm \frac{\tilde{\xi}_{\boldsymbol{k}}}{E_{\boldsymbol{k}}}\right). \label{eqn:wplusminus}
\end{align}
The energy dispersion $\xi_{\boldsymbol{k}} = - 2t(\cos
k_xa  + \cos k_ya ) - 4t^{\prime} \cos k_xa \cos k_ya - 2t''(\cos
2k_xa  + \cos 2k_ya )-\mu_p$ is the third
nearest-neighbor tight-binding energy dispersion, while $\xi_{\boldsymbol{k}}^0  =  - 2t(\cos k_xa  +
\cos k_ya)$ is the first nearest-neighbor term, which effectively shifts the
placement of the pseudogap off the Fermi surface to an energy which coincides with the
AFBZ boundary, defined by the half-filling point of the tight-binding energy $\xi_{\boldsymbol{k}}^0$.  These energy dispersions contain doping dependent
coefficients: $t(x)=g_{t}(x)t_{0}+3g_{s}(x)J\chi/8$, $t^{\prime
}(x)=g_{t}(x)t_{0}^{\prime }$, and $t^{\prime\prime
}(x)=g_{t}(x)t_{0}^{\prime \prime }$,  where $g_{t}(x)=\frac{2x}{1+x}$ and $g_{s}(x)=\frac{4}{(1+x)^{2}}$ are the energy renormalizing Gutzwiller factors for the kinetic and spin terms, respectively.  $g_t (x)$ also appears in Eq.~(\ref{eqn:sc}) as a weighting factor for the coherent part of the Green's function which acts to statistically remove or project out doubly occupied states.\cite{vollhardt:1984,gutzwiller:1963}
The dispersion, $\xi_{\boldsymbol{k}}$, uses $\mu_p$ as a chemical potential determined by the Luttinger sum rule.\cite{luttinger:1960} Values of
other parameters in the dispersion were taken from Ref.~\cite{yrz:2006} to be: $t^{\prime}/t_{0}=-0.3$, $t^{\prime\prime}/t_{0}=0.2$,
$J/t_{0}=1/3$, and $\chi=0.338$ which are accepted values for the hole doped cuprates \cite{ogata:2008}. The optimal superconducting gap
$\Delta_{\rm sc}^{0}$ was chosen to give an optimal $T_{c}$
around 95K for a ratio of $2\Delta^0_{\rm sc}(x,T=0)/k_{\rm
  B}T_c=6$, using of $t_{0}=175$ meV.  We also take $\Delta^0_{\rm sc}(x,T)$, the gap amplitude at doping, $x$, and temperature, $T$, to have a BCS temperature dependence.  We will assume that the pseudogap is only a function of doping, and hold the value of $\Delta_{\rm pg}(x)$ constant with temperature.  This will allow us to attribute any temperature dependence in calculated quantities as being due to the superconducting state.
$\Delta_{\rm sc}^0(x)$ and $\Delta_{\rm pg}^0(x)$ are described by the well known superconducting dome and pseudogap line, the latter vanishing at $T=0$ at a QCP in this model.  These are given explicitly as
\begin{align}
\Delta_{sc}^0(x,T=0)&= 0.14t_0(1-82.6(x-0.16)^2), \label{eqn:scdome}\\
\Delta_{pg}^0(x)&= 3t_0(0.2-x). \label{eqn:pgline}
\end{align}
\indent From the YRZ Green's function of Eq.~(\ref{eqn:sc}) and standard equations of superconductivity for the anomalous propogator, one can extract the regular and anomalous spectral functions, $A(\boldsymbol{k},\omega)$ and $B(\boldsymbol{k},\omega)$ respectively, and see that there are four energy branches, given by the energies, 
\begin{equation}
\pm E^{\alpha}_{\rm s}=\pm \sqrt{(E_{\boldsymbol{k}}^{ \alpha })^2  +
  \Delta _{\rm sc}^2 }.
  \end{equation}
  These energy branches appear in the spectral functions as,
  \begin{align}
  A(\boldsymbol{k},\omega)&=\sum_{\alpha=\pm} g_tW_{\boldsymbol{k}}^\alpha [(u^\alpha)^2\delta(\omega-E_{\rm s}^\alpha)+(v^\alpha)^2\delta(\omega+E_{\rm s}^\alpha)],\\
  B(\boldsymbol{k},\omega)&=\sum_{\alpha=\pm}  g_tW_{\boldsymbol{k}}^\alpha u^\alpha v^\alpha[ \delta(\omega+E_{\rm s}^\alpha) -\delta(\omega-E_{\rm s}^\alpha)],
  \end{align}
    where
    \begin{align}
 u^\alpha &=\left[\frac{1}{2}\left(1+\frac{E_{\boldsymbol{k}}^\alpha}{E_{\rm s}^\alpha}\right)\right]^{1/2},\\
 v^\alpha &=\left[\frac{1}{2}\left(1-\frac{E_{\boldsymbol{k}}^\alpha}{E_{\rm s}^\alpha}\right)\right]^{1/2}.
  \end{align}
   The Raman response, $\chi_{\eta}^{\prime\prime}(\Omega)=Im[ \chi_{\eta}(T,\nu)]$, is given by
\begin{eqnarray}
\chi^{\prime \prime}_{\eta}(\Omega)=\frac{\pi}{4}\sum_{\boldsymbol{k}}(\gamma_{\boldsymbol{k}}^\eta)^2\int_{-\infty}^{\infty}d\omega[f(\omega)-f(\omega+\Omega)]  \nonumber \\
\times [A(\boldsymbol{k},\omega)A(\boldsymbol{k},\omega+\Omega)+B(\boldsymbol{k},\omega)B(\boldsymbol{k},\omega+\Omega)], \label{eq:raman} 
\end{eqnarray}
where $\eta$ is the choice of vertex B$_{1g}$ or B$_{2g}$. The vertex strength, $(\gamma^\eta_{\boldsymbol{k}})^2$, can be determined straightforwardly from the energy dispersion, such that
\begin{align}
\gamma_{\boldsymbol{k}}^{B_{1g}}&=\frac{\partial^2 \xi(\boldsymbol{k})}{\partial k_x^2}-\frac{\partial^2 \xi(\boldsymbol{k})}{\partial k_y^2}, \\
\gamma_{\boldsymbol{k}}^{B_{2g}}&=\frac{\partial^2 \xi(\boldsymbol{k})}{\partial k_x \partial k_y}.
\end{align}
  For the energy dispersion used, this results in
\begin{align}
\gamma_{\boldsymbol{k}}^{B_{1g}}&=2ta^2(\cos k_x a- \cos k_y a)\nonumber \\
&+8t^{\prime\prime}a^2(\cos 2k_x a-\cos 2 k_y a), \label{eq:B1gvertex}\\
\gamma_{\boldsymbol{k}}^{B_{2g}}&=-4t^\prime a^2 \sin k_x a \sin k_y a.
\end{align}
\indent  The Raman response can be simplified by performing the integration over $\omega$ in Eq.~(\ref{eq:raman}).  This allows us to separate the contribution to the Raman spectra from each energy branch so that
\begin{align}
&\chi^{\prime \prime}_{\eta}(\Omega)=\frac{\pi}{4}\sum_{\boldsymbol{k},\alpha=\pm}(\gamma_{\boldsymbol{k}}^\eta)^2g_t\big(W^\alpha_{\boldsymbol{k}}[f(E_s^\alpha)-f(E_s^\alpha+\Omega)]  \nonumber \\
&\times \left[(u^{\alpha})^2 A(\boldsymbol{k},E_s^\alpha +\Omega) + u^\alpha v^\alpha B(\boldsymbol{k},E_s^\alpha +\Omega)\right] \nonumber\\ 
&+W^\alpha_{\boldsymbol{k}}[f(-E_s^\alpha)-f(-E_s^\alpha +\Omega)]\nonumber\\
&\times \left[(v^{\alpha})^2 A(\boldsymbol{k},-E_s^\alpha +\Omega) - u^\alpha v^\alpha B(\boldsymbol{k},-E_s^\alpha +\Omega)\right]\big).\label{eqn:ramangood}
\end{align}
    In the clean limit, Eq.~(\ref{eqn:ramangood}) can be written in the form
  \begin{align}
  &\chi^{\prime \prime}_{\eta}(\Omega)=\frac{\pi}{4}\sum_{\boldsymbol{k}}(\gamma_{\boldsymbol{k}}^\eta)^2g^2_t\Big([f(E_s^+)-f(E_s^-)]W^+_{\boldsymbol{k}}W^-_{\boldsymbol{k}} \nonumber \\
  &\times[u^+u^--v^+v^-]^2[\delta(\Omega+E_s^+-E_s^-)-\delta(\omega+E_s^--E_s^+)] \nonumber\\
  &+ [f(-E_s^+)-f(E_s^-)]W^+_{\boldsymbol{k}}W^-_{\boldsymbol{k}}\nonumber\\
  &\times[v^+u^-+v^-u^+]^2\delta(\Omega-E_s^+-E_s^-) \nonumber \\
&+   [f(-E_s^+)-f(E_s^+)](W^+_{\boldsymbol{k}})^2[2(u^+v^+)^2]\delta(\Omega-2E_s^+) \nonumber \\
&+   [f(-E_s^-)-f(E_s^-)](W^-_{\boldsymbol{k}})^2[2(u^-v^-)^2]\delta(\Omega-2E_s^-) \nonumber \\
&  +\Big\{h(E_s^+)(W^+_{\boldsymbol{k}})^2[(u^+)^2-(v^+)^2]^2   \nonumber \\
&\left. \hspace{12pt} +h(E_s^-)(W^-_{\boldsymbol{k}})^2[(u^-)^2-(v^-)^2]^2\Big\}\Omega\delta(\Omega) \right),\label{eqn:ramanbad}
  \end{align}
  
 \noindent where the function $h(E)=-\frac{\partial f(E)}{\partial E}$.  In this form one can see the interband (first and second terms), intraband (third and fourth terms) and Drude (last term which vanishes in the clean limit) contributions to the Raman spectra.  However, with moderate impurities Eq.~(\ref{eqn:ramanbad}) becomes inaccurate at low frequencies, due to not properly capturing the Drude contribution.  For this reason, throughout this work, all Raman calculations shown are the results of Eq.~(\ref{eqn:ramangood}).
  \section{Arpes}
In this section, we wish to understand how the angular energy profiles seen in ARPES data come about in the YRZ model.  This will be essential to the Raman analysis and discussion given in Sections IV and V. 

The concept of a Fermi surface, defined here by $\xi_{\boldsymbol{k}}=0$, can be a powerful aid in understanding the $k$-space distribution of electronic states and excitations.  In the presence of energy gaps, more complicated energy dispersions, $E_s^{\pm}$, arise which may not trace a zero energy surface.  In these cases, the examination of the minima of the energy dispersion produces a surface which represents the lowest energy excitations in $k$-space.
\begin{figure}
\centering
\includegraphics[width=0.95\linewidth]{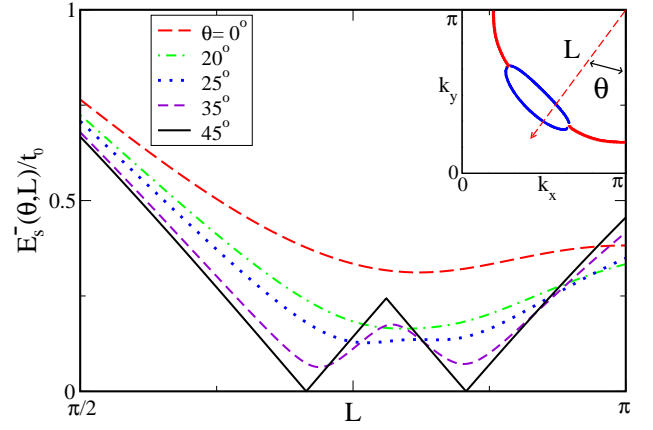}
\caption{\label{fig:slice}(Color online)  Plot of $E_s^-$ band as a function of length, L, along a line traced from the ($\pi$,$\pi$) (see inset) point for a range of angles, $\theta$, in the $x=0.14$ doping case.  The local minima in this curve are the energies of nearest approach, $\Delta_{na}$.  The inset shows the $k$-space locations of these nearest approach energies, marking out the Fermi pocket and its extension to the Brillouin zone boundary through the Fermi-Liquid-like `wings', shown in red.}
\end{figure}
In the YRZ model, examination of the $E_s^+$ and $E_s^-$ energies as a function of angle, serve to produce these lowest energy surfaces, or nearest approach energy, $\Delta_{na}$, surfaces.  In Fig.~\ref{fig:slice}, the $E_s^-$ energy is plotted as a function of distance along a line traced from ($\pi$,$\pi$) for a range of angles.  At $\theta=45^\circ$, the line is traced through the nodal direction, where both the pseudogap and superconducting gap equal zero.  This results in a true $E_s^-=0$ Fermi surface in the ($\pi$,$\pi$) direction, culminating at well known Dirac points.  For angles away from the node, there are no $E_s^-=0$ points.  It is for these angles that the minima in energy serves as an effective Fermi surface.

When these nearest approach energies are traced in $k$-space one obtains the inset of Fig.~\ref{fig:slice} demonstrating the Fermi pockets which are an essential part of the YRZ model.
\begin{figure}
 \centering
\subfigure{\includegraphics[clip,trim= 27mm 5mm 20mm 15mm,width=0.49\linewidth]{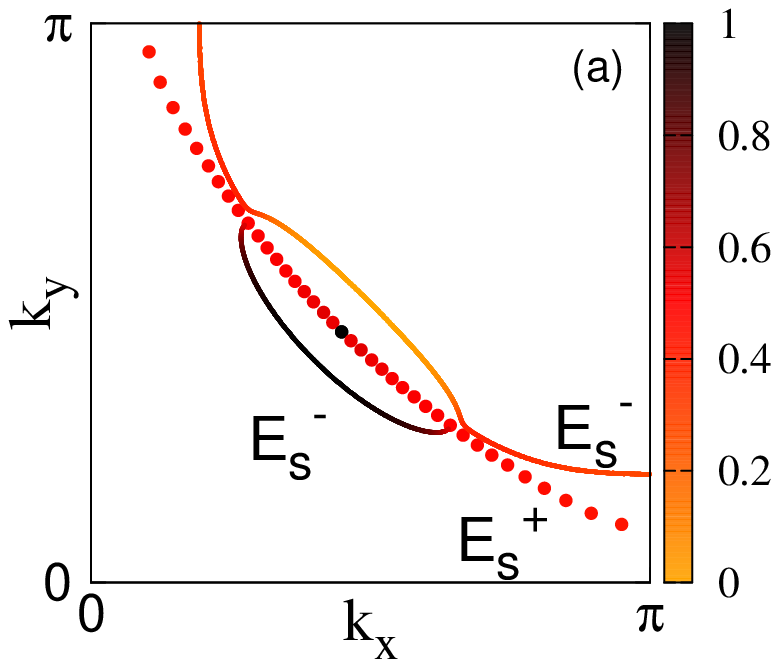}\label{fig:nasurface:a}}
\subfigure{\includegraphics[clip,trim= 27mm 5mm 20mm 15mm,width=0.49\linewidth]{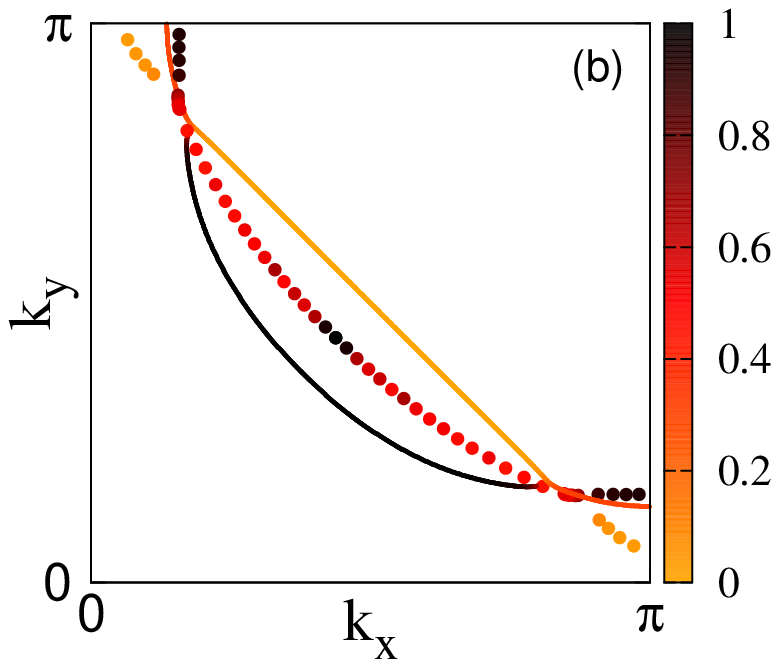}\label{fig:nasurface:b}} 
\caption{\label{fig:nasurface}(Color online) Locations of nearest approach for both the $E_s^+$ (dots) and $E_s^-$ (solid) bands for the (a) $x$=0.12 and (b) $x$=0.19 cases with the color scale of the individual pseudogap weightings $W^+_{\boldsymbol{k}}$ and $W^-_{\boldsymbol{k}}$.  This illustrates the restructuring of the tight-binding Fermi surface in the antinodal direction, shown as the black dots of the curve in (b), related to the $E_s^+$ band.  }
\end{figure}
As has been previously described, the $E_s^+$ and $E_s^-$ bands have relative weightings $W^+_{\boldsymbol{k}}$ and $W^-_{\boldsymbol{k}}$, given in Eq.~(\ref{eqn:wplusminus}) which are dependent upon both the sign of the bare dispersion, as well as modified by the presence of the pseudogap, such that $W_{\boldsymbol{k}}^++W_{\boldsymbol{k}}^-=1$.
These weightings are incorporated into the nearest approach momentum contours of Fig.~\ref{fig:nasurface} through a color scale as shown by two vertical columns with black equal to one and orange equal to zero.

Of course, each branch will have, possibly, separate nearest approach surfaces.  Examples of these are shown in Fig.~\ref{fig:nasurface}.  In Fig.~\ref{fig:nasurface:a}, which contains a large pseudogap, the $E_s^-$ band, shown as a continuous curve, dominates in weighting along the frontside of the Fermi pocket closest to (0,0), forming a strong Fermi arc, shown in black (the top of our color scale).  The backside of the Fermi pocket closest to the AFBZ boundary has a small weighting due to proximity to the pseudogap located there and is shown in orange which is the bottom of our color scale.  In a recent ARPES experiment \cite{meng:2009} the backside of the pocket has finally been resolved and is indeed seen with much lower intensity as compared with the front side.  The second momentum space contour, shown as solid dots, in Fig.~\ref{fig:nasurface:a} are associated with the $E_s^+$ branch.  For most of this curve the intensity falls in the middle of the range shown, which is red on our color scale.  Note that while these nearest approach contours are well defined, Fig.~\ref{fig:nasurface} tells us nothing about how close to zero the approach energies might be.  This can be very important in determining how effective a process involving such energies might be.  Later we will return to this issue.

Fig.~\ref{fig:nasurface:b} shows how the closest approach surfaces change as the pseudogap becomes small.  Here for doping $x=0.19$, which is only slightly less than $x_c=0.20$, the $E_s^+$ band has substantial weight near the antinodal `wings', shown as black dots.  One can see the reformation of the tight-binding Fermi surface as the combination of $E_s^+$ and $E_s^-$ black regions.  The side of the pocket nearest to the AFBZ boundary now has a more uniform weighting, approaching zero in value for $x=x_c$.  The contours also take on the shape of the AFBZ with which they almost overlap.  In the antinodal direction the $E_s^+$ band has two nearest approach values as does $E_s^-$ in the Fermi pocket region.

\begin{figure}
 \centering
\includegraphics[width=0.75\linewidth]{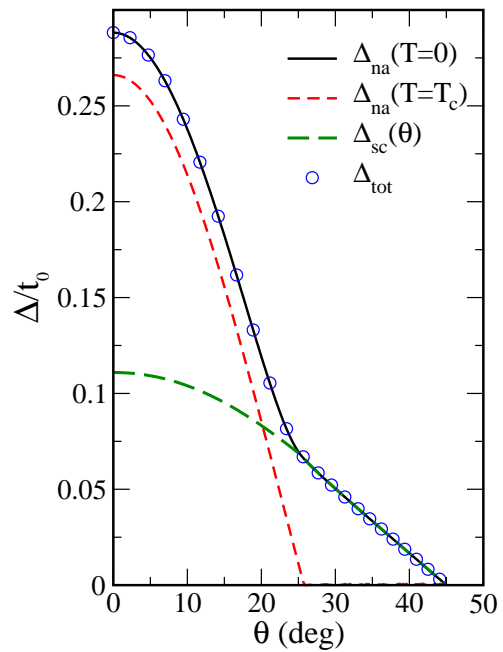}
\caption{ \label{fig:arpes1}(Color online)  $E_s^-$ energies of nearest approach, $\Delta_{na}$, versus angle, $\theta$, for the $x=0.12$ doping case: with superconductivity (solid black line) and without superconductivity (short-dashed red line).  The long-dashed green line is $\Delta_{\rm sc}(\theta)$ as described in the text.  Marked with open circles is the robust approximation that $\Delta_{na}(T=0,\theta)\approx\Delta_{tot}$, where $\Delta_{tot}(\theta)=\sqrt{\Delta_{\rm sc}^2(\theta)+\Delta_{na}^2(T_c,\theta)}$. The angle at which $\Delta_{na}(T_c)=0$ marks the edge of the Fermi pocket and thus defines $\theta_{yrz}$, as measured from $(\pi,\pi)$ and described in the text.}
\end{figure}

We can also examine the size of the nearest approach energies as a function of angle, along the strongly weighted region of Fig.~\ref{fig:nasurface:a}.  This is shown in Fig.~\ref{fig:arpes1}, where we compare the cases with either a pseudogap or a superconducting gap to the case with both.  
The surface traced by minima in the $E_s^-$ band coincides with the Luttinger pocket and actual $E_s^-=0$ Fermi surface in the absence of superconductivity, which is important for correctly describing low energy nodal excitations.  The strongly weighted part of the $E_s^+$ band in the nodal direction occurs at a high energy relative to a zero energy Fermi surface that exists in the $E_s^-$ band in the ($\pi$,$\pi$) direction.  Comparison with ARPES data with energy scales relative to the Fermi level will therefore be dominated by the $E_s^-$ band.  It is an important note that the nearest approach contours are not shifted when the superconductivity is switched on or off.  We make two calculations of $E_s^-(\theta)$: one without the superconductivity in which we denote the energy of nearest approach by $\Delta_{na}(T=T_c)$ (red short-dashed curve) and one with superconductivity denoted by $\Delta_{na}(T=0)$ (solid black curve).  We can also evaluate directly the size of the superconducting gap on the nearest approach contour, $\Delta_{\rm sc}(\theta)$ and this is the long-dashed green curve.  In the pseudogap case, for angles less than some critical angle, $\theta_{yrz}$, $\Delta_{na}(T=T_c)$, which contains only a pseudogap, is finite.  These angles $\theta <\theta_{yrz}$ represent the region along the `wings' of the Fermi surface.   In this sense $\theta_{yrz}$ marks the corner of the Fermi pocket.  For $\theta>\theta_{yrz}$, $\Delta_{na}(T=T_c)$ is uniformly zero along the strongly weighted side of the Fermi pocket.   In the YRZ model, in spite of introducing a d-wave pseudogap over the entire Brillouin zone the net effect is that it only appears dominant in the wings, and not in the pocket itself where the excitation spectrum shows no gap, as is also seen in ARPES data.  Fig.~\ref{fig:arpes1} also includes $\Delta_{na}(T=0,\theta)$ (solid black curve) which has both gaps present.  The value of $\Delta_{na}(T=0,\theta)$ is dominated at $\theta<\theta_{yrz}$ by the pseudogap contribution, and by the superconducting contribution for $\theta>\theta_{yrz}$.  However, there appears here a natural energy scale which is well approximated by $\Delta_{na}(T=0,\theta)\approx\Delta_{tot}(\theta)=\sqrt{\Delta_{\rm sc}^2(\theta)+\Delta_{na}^2(T_c,\theta)}$, the root sum of squares of the superconducting and pseudogap contributions.  This square root is represented by the open blue circles which overlap precisely with the $\Delta_{na}(T=0)$ curve.

\begin{figure}
 \centering
\includegraphics[width=0.75\linewidth]{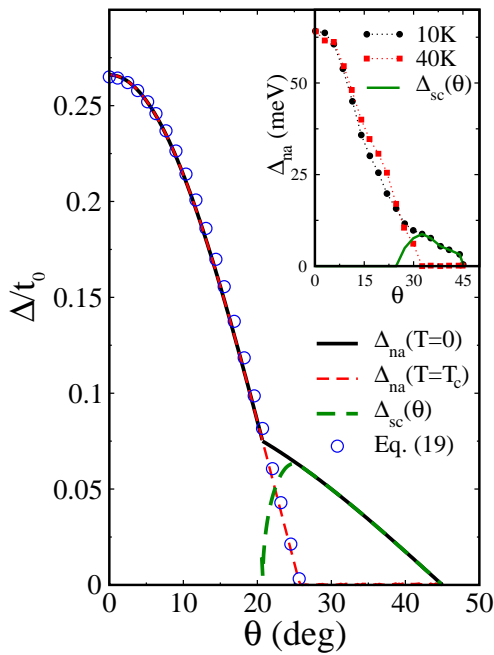}
\caption{\label{fig:arpes2} (Color online)  Nearest approach energies as a function of angle, $\theta$ [measured from $(\pi,\pi)$], at $T=0$ (solid black line) and $T=T_c$ (dashed red line) for doping $x=0.12$.  Here the superconducting gap profile $\Delta_{\rm sc}(\theta)$ (long-dashed green line) is chosen so that $\Delta_{na}(T=0,\theta)\approx\Delta_{na}(T=T_c,\theta)$ for small $\theta$, which is different from Fig.~\ref{fig:arpes1}, and is governed by $\Delta_{sc}(\theta) = \sqrt{\Delta_{na}^2(0,\theta) - \Delta_{na}^2(T_c,\theta)}$.  Inset shows a similar analysis of experimental data from Kondo \emph{et al} \cite{kondo:2009} for an underdoped sample measured at $T=10$K (low temperature) and $T=40$K (above $T_c$), where we take $\Delta_{sc}(\theta)=\sqrt{\Delta_{na}^2(10 \rm{K},\theta) - \Delta_{\mit na}^2(\rm 40 \rm{K},\theta)}$. }
\end{figure}

Comparison with experiment \cite{kondo:2009} yields a complication in that there appears to be minimal difference between $\Delta_{na}(T_c,\theta)$ and $\Delta_{na}(T=0,\theta)$ for small angles.  In order to obtain a similar feature that $\Delta_{na}(0,\theta)\approx\Delta_{na}(T_c,\theta)$ for small angles, corresponding to the region off of the Fermi pocket, one would infer from the equation for $\Delta_{tot}(\theta)$ that the new superconducting gap should be zero off of the Fermi pocket and be equal to $\Delta_{\rm sc}(\theta)$ from Fig.~\ref{fig:arpes1} for large angles, corresponding to the region on the Fermi pocket.
Guided by Fig.~\ref{fig:arpes1}, we have chosen a $\Delta_{\rm sc}(\theta)$ which is zero for small angles and smoothly transitions to be equivalent to its value in Fig.~\ref{fig:arpes1} for large angles.  Fig.~\ref{fig:arpes2} demonstrates the result of such a constraint on $\Delta_{\rm sc}(\theta)$ (long-dashed green curve) as well as the expected overall gap $\Delta_{na}(0,\theta) = \sqrt{\Delta_{sc}^2(\theta) + \Delta_{na}^2(T_c,\theta)}$ (solid black).  This procedure results in a superconducting gap
profile which appears either highly nonmonotonic, or that is only present on the Fermi pocket for $\theta>\theta_{yrz}$, dropping off quickly beyond that.  One can apply this simple analysis to the experimental data taken from Ref.~\cite{kondo:2009} which we have shown in the inset of Fig.~\ref{fig:arpes2}.  We see that the superconducting gap (solid green curve) is nonzero only in the nodal region up to approximately 23$^\circ$, \emph{i.e.} between 23$^\circ$--45$^\circ$

The issue of how the superconducting gap presents itself on the Fermi surface in complicated electronic systems is important.  Even in conventional metals, for which the electron-phonon interaction is responsible for the superconducting condensation, the gap is found to be highly anisotropic and does not exist for certain solid angles in momentum space where there is no Fermi surface because it is gapped out by the crystal potential.\cite{tomlinson:1976,leung:CJP:1976,leung:1976}  In the cuprates, significant deviations from the simplest $d$-wave gap, Eq.~(\ref{eqn:scgap}), have been known for some time in Bi$_2$Sr$_2$CaCu$_2$O$_{8+\delta}$\cite{mesot:1999} and are seen in ARPES data which resolves bilayer splitting in (Pb,Bi)$_2$Sn$_2$CaCuO$_{8+\delta}$.\cite{borisenko:2002}  A nonmonotonic gap was also observed in Nd$_{2-x}$Ce$_x$CuO$_4$,\cite{blumberg:2002} an electron doped superconductor.  This is also expected in theoretical models with pairing based on spin fluctuations.\cite{o'donovan:1995:4568,o'donovan:1995:16208,o'donovan:1996, branch:1995}  Numerical solutions of the gap equation in such cases show that many higher $d$-wave harmonics are present in the gap function.  When a pseudogap is present, as is the case in the underdoped cuprates, one expects it to prevent the lowest order harmonic contribution to the superconducting gap from having its full amplitude.  This amplitude could even be forced to zero as is indicated in the data shown in the inset of Fig.~\ref{fig:arpes2} for the regions away from the Fermi pocket.  The Fermi pocket is the only region where the normal state supports zero energy excitations.

Of additional note, the angular profile of the pseudogap, $\Delta_{na}(T_c,\theta)$ can be compared to another model, such as the arc model, which has previously been used to fit electronic specific heat data \cite{storey:2007,leblanc:2009} as well as understand the existence of two energy scales in Raman spectra.\cite{storey:CAP:2008}  The arc model, named as such because it produces a Fermi arc rather than a Fermi pocket, places a pseudogap over only a portion of the large tight-binding Fermi surface, \emph{i.e.},
\begin{equation}
\Delta^{\rm ARC}_{\rm pg}(\boldsymbol{k})=\left\{
\begin{array}{ll}\displaystyle
\Delta_{\rm pg}^{0}\cos\left(\frac{\pi \theta}{2\theta_o}\right),  &
(\theta<\theta_o)\\
\\
\displaystyle
\Delta_{\rm pg}^{0}\cos\left(\frac{\pi(\theta-\pi/2)}{2\theta_o}\right), &
(\theta>\frac{\pi}{2}-\theta_o).
\end{array}\right.
\label{eq:pgeqn}
\end{equation}
  This $\Delta_{\rm pg}^{\rm ARC}(\theta)$ is essentially a shrunken $\cos(2\theta)$ which goes to zero at $\theta_0$ rather than $\theta=45^\circ$.  For reference, the open blue circles in Fig.~\ref{fig:arpes2} show Eq.~(\ref{eq:pgeqn}) for $\theta_0=\theta_{yrz}$.  This arc model pseudogap profile shows minimal difference from what would be extracted on the heavily weighted side of the pocket in the YRZ model.  In this respect the two models are equivalent, the arc is just one side of the pocket.
 
\section{Raman in the YRZ model}
Experimental Raman spectra suffer from a loss of the B$_{1g}$ signal with increased underdoping as well as a strong signal from inelastic scattering at higher frequencies and complications due to surface effects.  Still, given these issues, work has been done which is able to see clearly two energy scales:\cite{guyard:2008,guyard:PRL:2008,tacon:2006} one in the antinodal B$_{1g}$ signal which increases and the other in the nodal B$_{2g}$ signal  which decreases with underdoping.
While ARPES experiments are able to resolve $k$-space dependence, Raman spectra sample a finite region of $k$-space associated with nodal or antinodal polarizations.  These regions may even overlap somewhat.  We seek to illustrate how some of the intricate details seen in ARPES, namely the Fermi surface restructuring and the superconducting gap profile, impact upon Raman spectra.

\begin{figure}
 \centering
\includegraphics[width=0.96\linewidth]{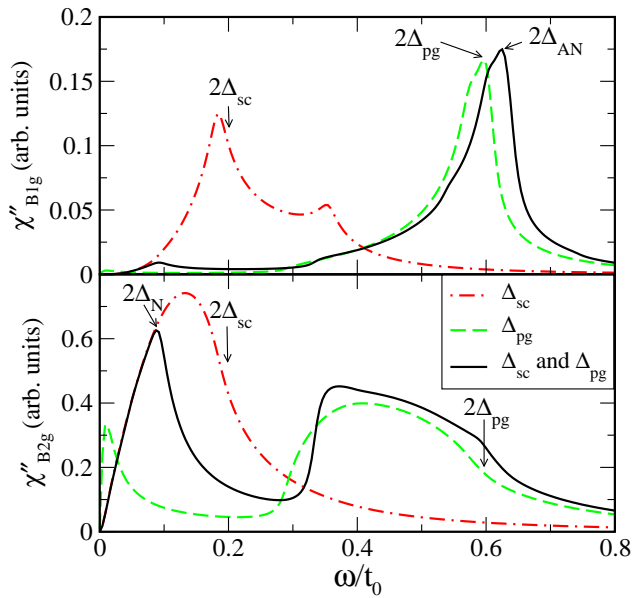}
\caption{ \label{fig:raman1}(Color online)  Raman response curves for the $x$=0.10 case.  The top and bottom frames show the B$_{1g}$ and B$_{2g}$ responses, respectively.  The antinodal gap, $2\Delta_{AN}$, and the nodal gap, $2\Delta_{N}$, are defined as the dominant peaks in their respective B$_{1g}$ and B$_{2g}$ curves.  As well, the actual input gap values $2\Delta_{sc}=0.2t_0$ and $2\Delta_{pg}=0.6t_0$ are marked.}
\end{figure}

Fig.~\ref{fig:raman1} shows a sample calculation for the $x$=0.10 case.  Each frame contains Raman spectra for the pseudogap case only (dashed green), superconducting case only (dash-dotted red) and both (solid black).  In the top frame the B$_{1g}$ peak, $2\Delta_{AN}$, is marked.  For this doping, the $2\Delta_{AN}$ peak occurs at a higher frequency than both the $2\Delta_{\rm pg}$ and $2\Delta_{\rm sc}$ points due to the additive impact of these two gaps on the antinodal gap.  In the $\Delta_{\rm sc}$ only curve, the peak occurs just before the $2\Delta_{\rm sc}$ point.
In the simplest BCS calculations based on free electron bands, the position of this peak would coincide exactly with the value of twice the superconducting gap amplitude.  However, in more complicated band structure models, as we are using here, and with a superconducting gap defined over the entire brillouin zone rather than just on the Fermi surface this need not be the case.  In addition, the Raman cross-section can display other features such as the second peak seen in the red dash-dotted curve of the top frame of Fig.~\ref{fig:raman1} which have their origin in the band structure.
In the bottom frame, the B$_{2g}$ peak, $2\Delta_{N}$ (first peak in the B$_{2g}$ spectrum), is marked on the solid black line, and occurs at lower energy than both $2\Delta_{\rm sc}$ and $2\Delta_{\rm pg}$.  There are two main factors which contribute to this lower value.  First, the value of a gap in the B$_{2g}$ spectra in standard BCS theory falls not at its peak but beyond the peak, at a point of inflection.  Additionally, since the B$_{2g}$ vertex samples largely the nodal direction, it may not contain the maximum input gap values which occur in the antinodal direction, depending upon the actual size and shape of the Fermi pocket.

\begin{figure}
 \centering
\includegraphics[clip,width=0.96\linewidth]{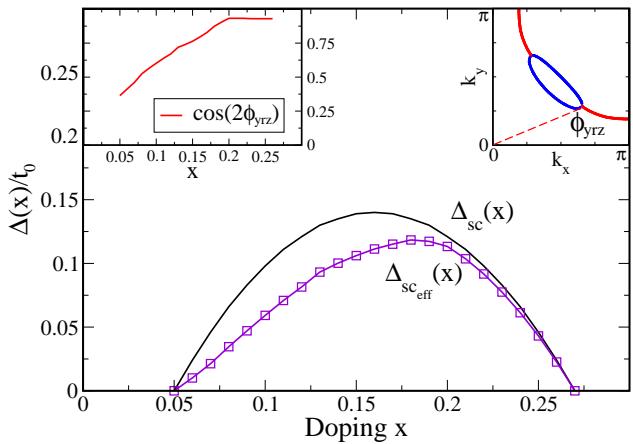}
\caption{\label{fig:sceff}(Color online)  Plot of the superconducting dome $\Delta_{\rm sc}(x)$ (black) and the reduced effective gap $\Delta_{\rm sc_{eff}}(x)=\Delta_{\rm sc}(x)\cos(2\phi_{yrz})$ (open purple squares).  This effective gap represents the maximum gap value observed on the Fermi pocket, which occurs at an angle $\phi_{yrz}$ as measured from (0,0) (see right inset).  The factor $\cos(2\phi_{yrz})$ is plotted as a function of doping, $x$, in the left inset. }
\end{figure}

To elaborate, as mentioned in Fig.~\ref{fig:arpes1}, the edge of the Fermi pocket is marked by an angle $\theta_{yrz}$ which changes quite drastically with doping.  If we instead measure this angle from (0,0), which we will distinguish by calling it $\phi_{yrz}$ (shown in the upper right inset of Fig.~\ref{fig:sceff}), we would find that $\phi_{yrz}$ has a large value approaching 45$^\circ$ at $x=0$, and for dopings at or above the QCP, $x_c$, has some small value defined by the angle from (0,0) to the point where the tight binding Fermi surface intersects the Brillouin zone boundary.  As shown in Fig.~\ref{fig:nasurface:a} the edge of this pocket can be quite heavily weighted, and as a result any experimental probes that are concerned with excitations along the Fermi surface will see a maximum superconducting gap value at the edge of the Fermi pocket near the angle $\phi_{yrz}$ where a maximum value coincides with a strong weighting.  
In a theoretical calculation, the input of a superconducting gap with a maximum value $\Delta_{\rm sc}^0(x)$ [see Eq.~(\ref{eqn:scgap})] such that $\Delta_{\rm sc}(x,\phi)=\Delta_{\rm sc}^0(x)\cos(2\phi)$ will show features in the Raman spectra at a value $\Delta_{\rm sc_{eff}}(x)=\Delta_{\rm sc}(x,\phi_{yrz})$.  Fig.~\ref{fig:sceff} demonstrates the size of this effect in the YRZ model.  Shown in solid black is the input superconducting dome given by Eq.~(\ref{eqn:scdome}) while the open purple squares show $\Delta_{\rm sc_{eff}}$ in the YRZ model.  This is obtained by finding the angles $\phi_{yrz}(x)$ across the doping phase diagram.  Although this effect is small in the absence of pseudogap; in the presence of pseudogap, this alone results in a factor of 2 difference between the apparent gaps, as compared with the actual input gap, across the phase diagram.  This is seen clearly in the upper left inset of Fig.~\ref{fig:sceff}, which shows just the value of $\cos(2\phi_{yrz})$ versus doping.  

\begin{figure}
  \centering
  \subfigure{\label{fig:ud}\includegraphics[clip,trim= 27mm 5mm 22mm 15mm, width=0.6\linewidth]{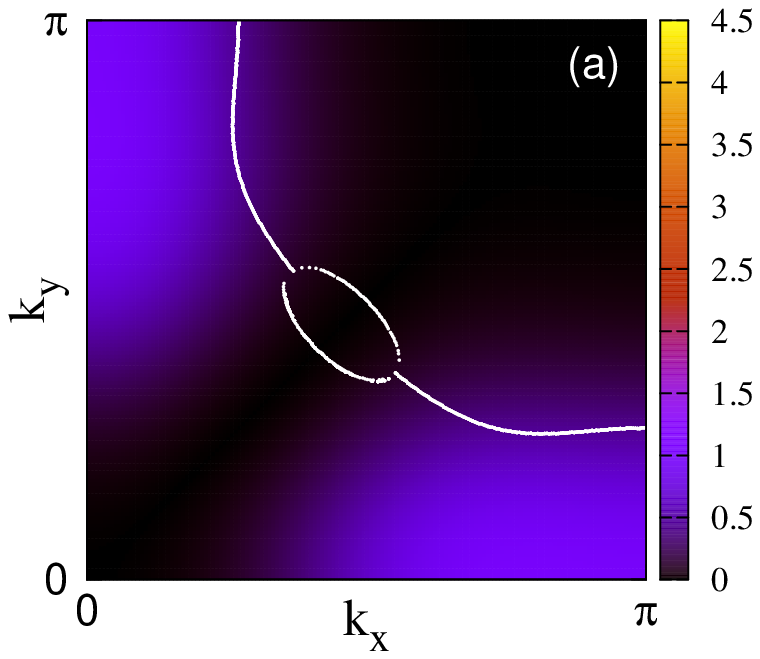}}      \\          
  \subfigure{\label{fig:opt}\includegraphics[clip,trim= 27mm 5mm 22mm 15mm, width=0.6\linewidth]{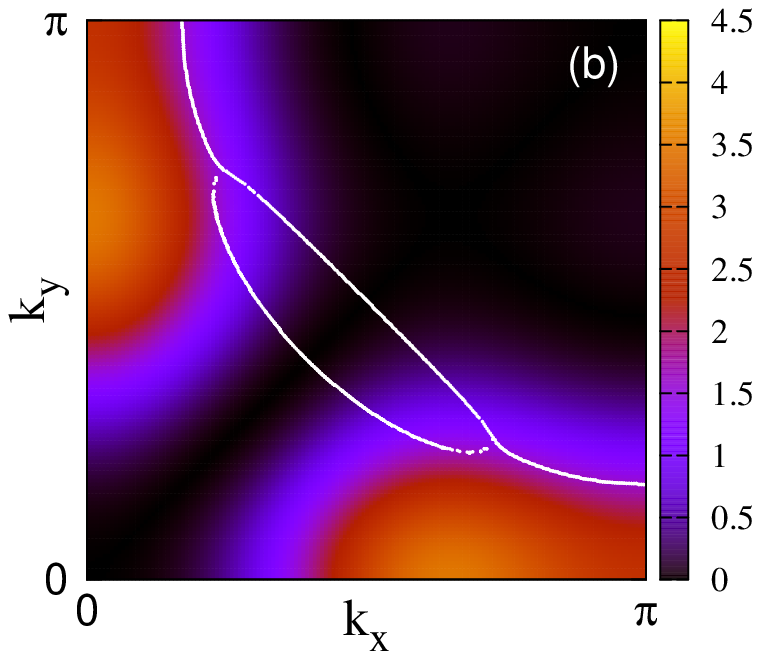}} \\
  \subfigure{\label{fig:od}\includegraphics[clip,trim= 27mm 5mm 22mm 15mm, width=0.6\linewidth]{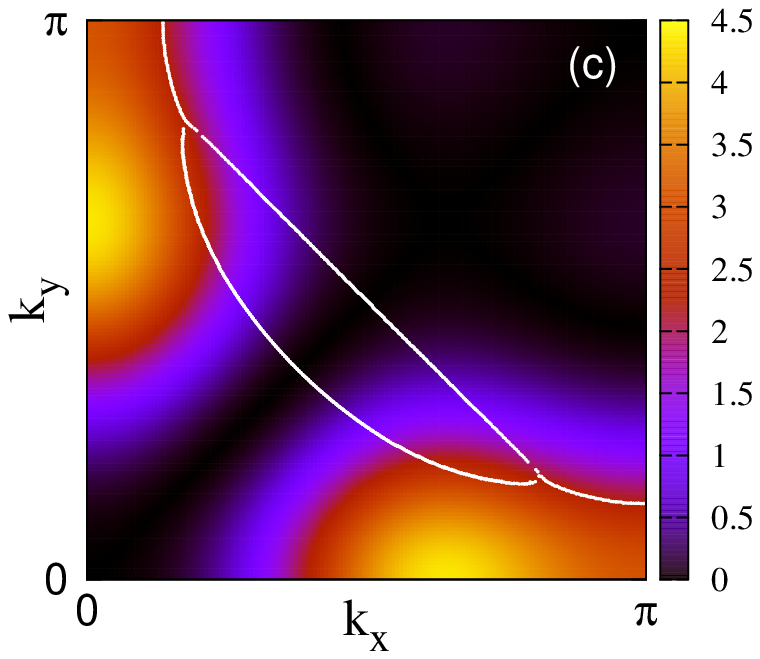}} 
  \caption{(Color online)  An overlay of the Fermi pocket and Fermi surface wings (in white) for the $E_s^-$ band, on a backdrop of the B$_{1g}$ vertex strength, $(\gamma^{B_{1g}})^2$, as in Eq.~(\ref{eq:B1gvertex}), for dopings: (a) $x$=0.05  (b) $x$=0.16 and (c) $x$=0.19.  This illustrates an expected loss of B$_{1g}$ signal in the underdoped region as the pocket, which dominates the superconducting state, shrinks away from regions of strong $(\gamma^{B1g})^2$, in addition to loss associated with the considerable weakening of the vertex itself.}
  \label{fig:2dplots}
\end{figure}

\begin{figure}
 \centering
\includegraphics[width=0.9\linewidth]{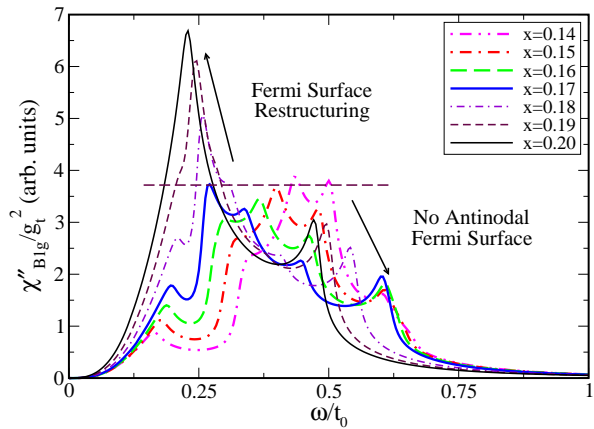}
\caption{(Color online)  Raman B$_{1g}$ spectra versus frequency for a range of doping values wherein the $g_t^2(x)$ prefactor has been removed.  Below $x$=0.17, the $\chi^{\prime\prime}_{B_{1g}}$ amplitude is small at low frequencies but maintains a roughly constant peak amplitude.  Above $x$=0.17, we see a strong increase in low frequency amplitude caused by the reformation of Fermi surface in the antinodal region which is of interest for the B$_{1g}$ vertex.  A dashed horizontal line has been added to help the reader distinguish these two regions.}\label{fig:restruct}
\end{figure}

The shrinking of the Fermi pocket with underdoping has additional important effects on the B$_{1g}$ antinodal response.  Fig.~\ref{fig:2dplots} shows an overlay of the effective Fermi surface onto the B$_{1g}$ vertex strength, $(\gamma_{\boldsymbol{k}}^{B_{1g}})^2$.  The B$_{1g}$ vertex cannot properly sample the Fermi surface near the edge of the Fermi pocket for strongly underdoped cases.  Although this is an extreme example, illustrating the large range of pocket size, this effect can be seen clearly over a much smaller doping range.  Fig.~\ref{fig:restruct} shows the YRZ Raman B$_{1g}$ spectra with the doping-dependent Gutzwiller factors removed to better illustrate this vertex strength effect.  For doping above $x=0.17$, there are extremely strong low frequency peaks which mark the antinodal gap.  For dopings below $x=0.17$ the low frequency signal has become extremely suppressed.  
The horizontal dashed line helps the reader to see the two distinct regions of doping with the boundary between the regions occuring at $x=0.17$.  
This doping falls considerably below its critical value, $x_c$=0.20, which defines the quantum critical point at which the pseudogap sets in.  In both regimes, the Raman vertex favors the antinodal direction.  While it can vary somewhat, both in shape and in amplitude, since the hopping parameters in the electronic dispersion curves depend on $x$, this is not the important effect that we wish to emphasize here.  What is more important is the evolution in the topology of the Fermi surface, or more correctly the surface of closest approach, that occurs for $x<x_c$.  As the pockets begin to shrink they move away from the antinodal point and the intensity of the Raman signal consequently drops due to an effectively smaller value of the vertex projected onto the pocket.
To get a large reduction it is necessary that the distance in momentum space between the end of the pocket and the Brillouin zone boundary at the antinode is not too small.  Our calculations show that this occurs rather abruptly at a doping of $x=0.17$.

\begin{figure}
  \centering
\hspace*{18pt}\subfigure{\includegraphics[width=0.4\linewidth]{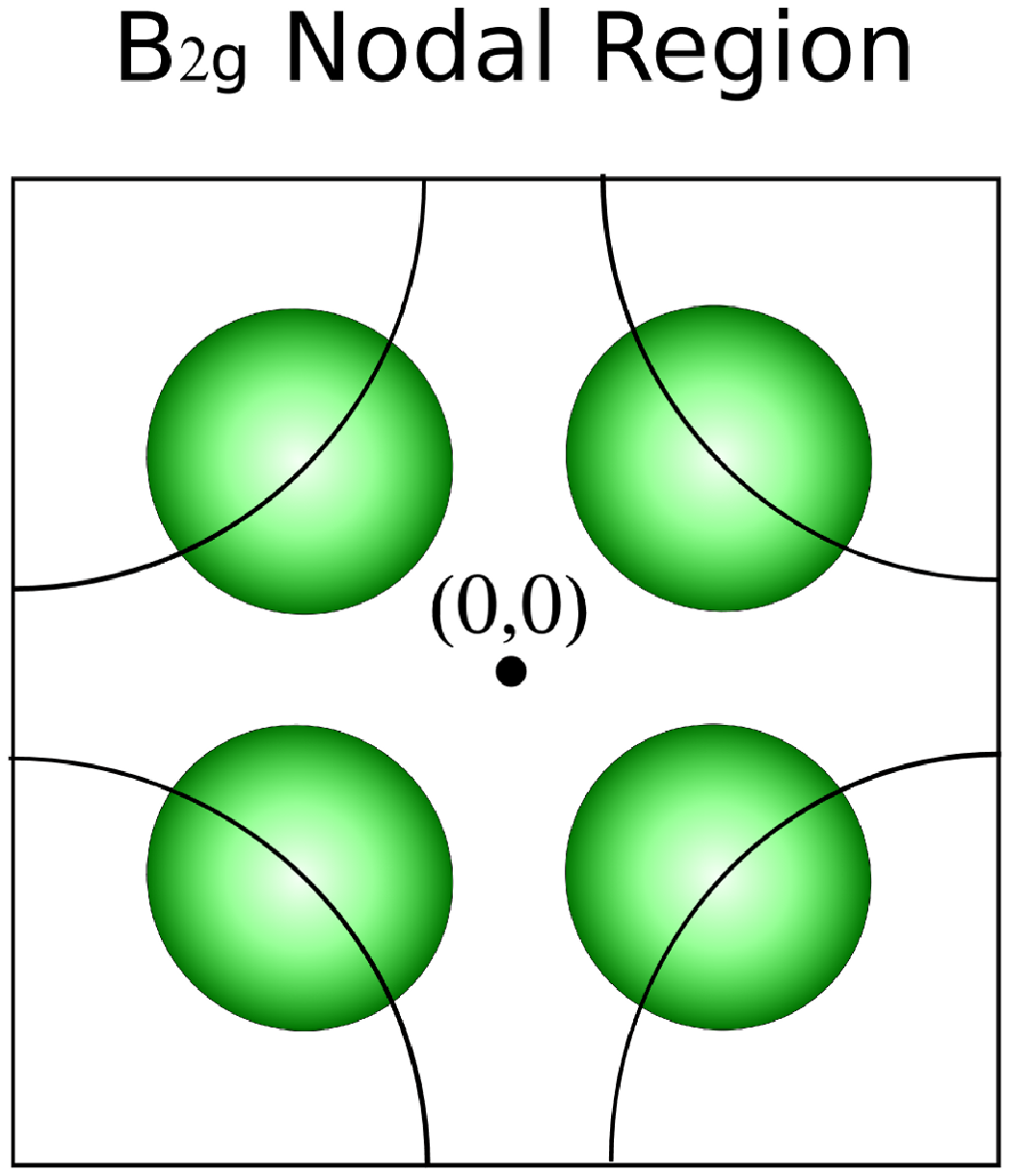}}%
\hspace*{18pt}\subfigure{\includegraphics[width=0.4\linewidth]{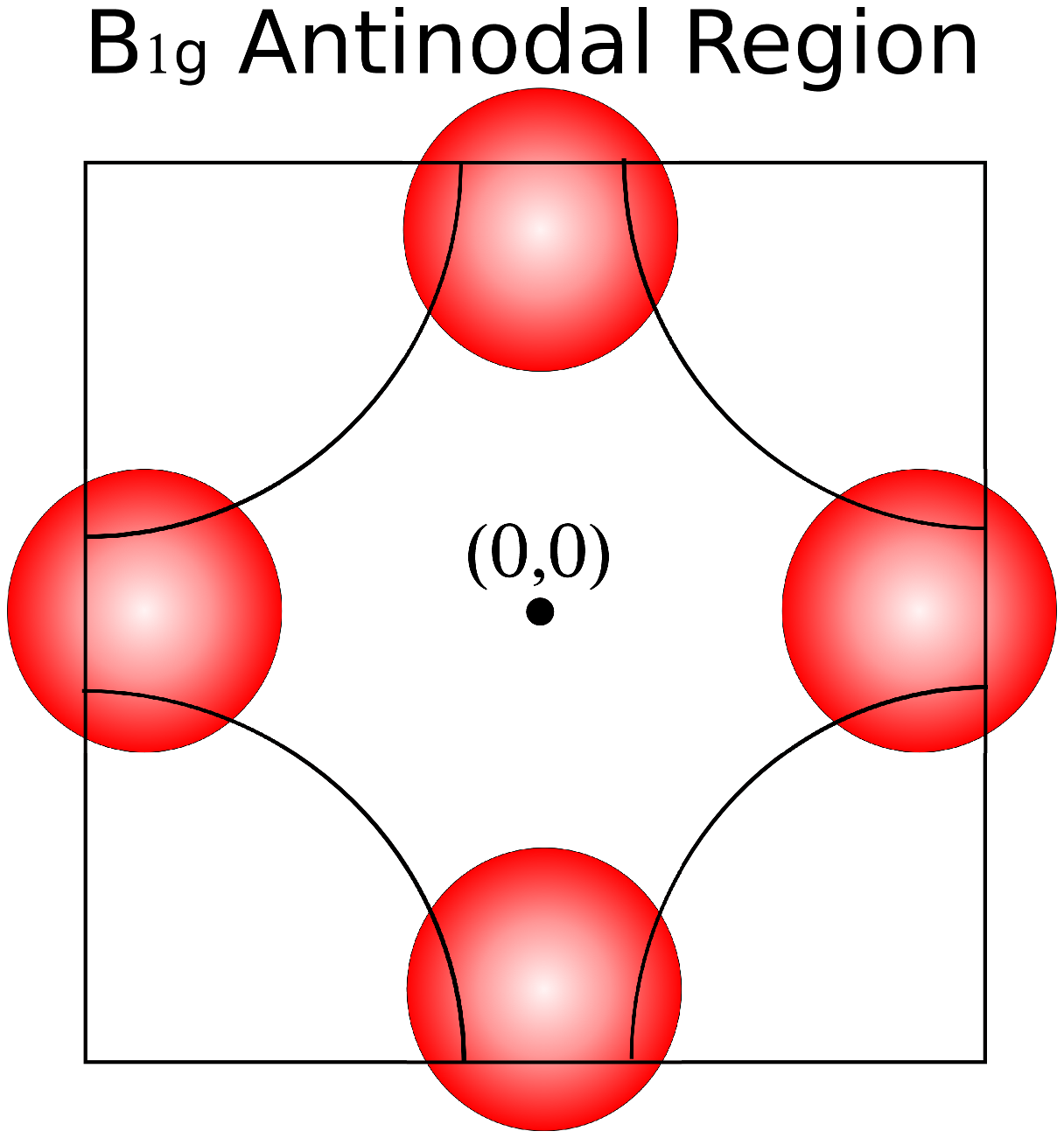}}\\%
\subfigure{\includegraphics[clip,height=.9\linewidth]{5graph-b2g-pure-shaded-scaled.eps}}%
\subfigure{\includegraphics[clip,height=.9\linewidth]{5graph-b1g-shaded.eps}}%
   \caption{ (Colour online) Raman response in the YRZ model for the B$_{2g}$ (left frames) and B$_{1g}$ (right frames) polarizations.  The dashed red curves are for the normal state, and act as a background for reference.  In the underdoped region the pseudogap background dominates, and superconductivity plays little role in the B$_{1g}$ spectra.  This is emphasized by the shaded region between the curves above the background peak.    The effect of $\Delta_{sc}$ on the B$_{2g}$ curves is clear, even at low dopings, by a strong B$_{2g}$ signal relative to the background.  Also note that the y-axis scales are modified with doping, which is an effect of the $g_t^2(x)$ prefactors.}
  \label{fig:5frames}
\end{figure}

So far we have not accounted for damping.  The inelastic scattering in the cuprates, which is known to be large, can broaden the Raman response considerably and provide a background that remains in the Fermi liquid state.\cite{jiang:1996,branch:2000}  In our formalism it would correspond to the incoherent part of the Green's function which should add a second piece in Eq.~(\ref{eqn:sc}) which involves only the coherent part.
It has been shown that the Raman, optical, and quasiparticle scattering rates are all strongly dependent on frequency.\cite{branch:2000,schachinger:2000,schachinger:2003,carbotte:2005}  In the superconducting state, these scattering rates are reduced at low energy by the opening of a superconducting gap and there is therefore no justification for broadening the B$_{2g}$ spectrum which samples mainly the low energy nodal region.  However, the antinodal B$_{1g}$ spectrum contains peaks that generally fall at higher energies, particularly in the underdoped region of the phase diagram, and we therefore expect rather large scattering rates and considerable broadening.
For experimental purposes, one would need to know the pseudogapped normal-state Raman response to compare to the superconducting Raman spectra in order to see indications of the strength of the superconducting state in the combined spectra.  Fig.~\ref{fig:5frames} gives such a comparison for the B$_{1g}$ (right frame) and B$_{2g}$ (left frame) spectra where the B$_{1g}$ curves have been broadened with a doping dependent elastic scattering of $\Gamma=0.2t(x)$ while the B$_{2g}$ have been kept near the clean limit. 
There is an apparent shift of B$_{1g}$ spectral weight, shown by the shaded yellow region, upon the addition of a superconducting energy scale.  The differences between the normal and superconducting B$_{1g}$ spectra become much less pronounced for increased underdoping.  This behaviour occurs at all dopings where the pseudogap energy scale dominates, and the superconductivity provides only a very small additional contribution to the antinodal Raman.  We note also that the overall strength of the Raman signal decreases with decreasing doping as noted in experiments.\cite{guyard:2008,tacon:2006}  As we have described before, this can be traced to a smaller overlap between the appropriate Raman vertex weighting factor in the Brillouin zone and the Fermi surface pocket.  
The case of the B$_{2g}$ response yields a different result.  The region of interest occurs at a lower frequency, on the scale of the superconducting energy gap, while the effect of the pseudogap is most apparent only at higher frequency.  However, the pseudogap does suppress the lower frequency normal-state Raman background.  This makes the effect of the superconducting gap very apparent relative to this background even at small dopings.  Note that in the lower left frame, the shaded yellow region at the lowest doping remains substantial compared with its value in the upper frame.  This is distinct from the case in the right hand frame where the shaded yellow region at low dopings has essentially disappeared.

\begin{figure}
 \centering
\includegraphics[width=0.96\linewidth]{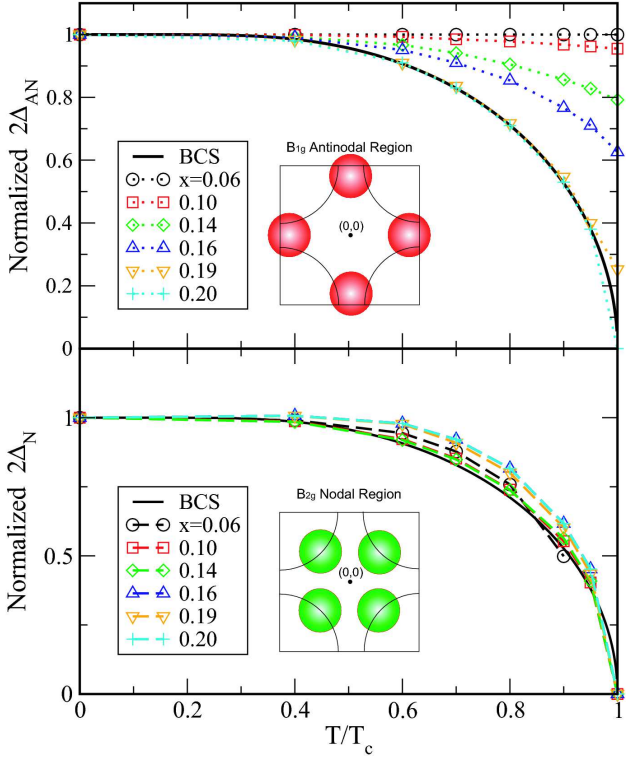}
\caption{ (Color online)  The normalized $\Delta_{AN}/\Delta_{AN}(T=0)$ (top frame) and $\Delta_{N}/\Delta_{N}(T=0)$ (bottom frame), as would be extracted from spectra similar to those in Fig.~\ref{fig:raman1} but taken at various finite temperature, are plotted versus $T/T_c$ for a range of dopings.  The nodal gap, $\Delta_N$, shows BCS-like temperature dependence across all dopings, while the $\Delta_{AN}$ diverges from BCS temperature dependence in the presence of a modest pseudogap.  Although not shown, this divergence from BCS behaviour is reasonably well-described by $\Delta(T)=\sqrt{\Delta_{\rm sc}^2(T)+\Delta_{\rm pg}^2}$.}\label{fig:temperature1}
\end{figure}

We have calculated the temperature evolution of these peaks for the case where the superconducting and pseudogaps are full $d$-wave gaps over the Brillouin zone, given by Eqs.~(\ref{eqn:scgap}) and (\ref{eqn:pggap}) and where we have assumed $\Delta_{\rm pg}^0$ to be constant with temperature and $\Delta_{\rm sc}^0$ to have a BCS temperature dependence.  The top frame of Fig.~\ref{fig:temperature1} shows that the B$_{1g}$ peaks, normalized by their zero temperature values, do not follow the input BCS temperature dependence for the superconducting gap, but follow instead a temperature dependence related to the combined magnitudes of $\Delta_{\rm sc}(T)$ and $\Delta_{\rm pg}$.  For the extremely underdoped cases, there is almost no discernible temperature dependence.  For increased doping, the pseudogap is reduced, and as $x_c$ is approached, the assumed BCS temperature dependence returns.
This is distinct from the temperature dependence in the nodal direction which is dominated by the temperature dependence of the superconducting gap as is shown in the bottom frame of Fig.~\ref{fig:temperature1} where it is compared to the input BCS $\Delta(T)/\Delta(0)$.

We also wish to consider a cutoff superconducting gap, a gap only present along the Fermi pocket, as suggested by ARPES experiments.  As a first attempt, we can cut off the superconducting gap sharply at the edge of the Fermi pocket.  Although this sharp cutoff would result in clear features in nearest approach energies, the sharpness should not create additional features in the Raman spectra due to the averaging effects of the Raman vertices.

\begin{figure}
 \centering
\includegraphics[width=0.96\linewidth]{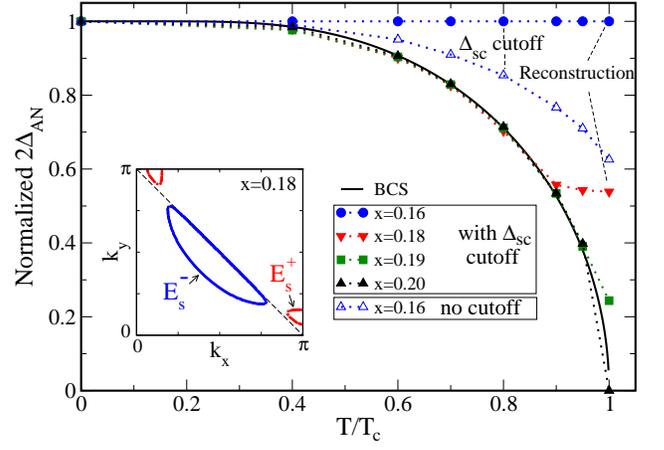}
\caption{ \label{fig:temperature2}(Color online) Normalized antinodal peak frequencies as a function of temperature.  Here, solid points illustrate the effect of cutting off superconductivity in the antinodal regions as in Eq.~(\ref{eq:sccut}).  This cutoff would result in no temperature dependence in the B$_{1g}$ peaks for any doping below $x$=0.16.  Above $x$=0.16, the Fermi surface begins to restructure in the antinodal directions and begins to be sampled by the B$_{1g}$ vertex.  Once this occurs, BCS-like temperature dependence returns.  The inset shows the reformation of the antinodal Fermi surface for the $x$=0.18 case.}
\end{figure}

To state explicitly, when we state `$\Delta_{\rm sc}$ cutoff', we are referring to
\begin{equation}
\Delta_{\rm sc}=\left\{
\begin{array}{ll}\displaystyle
\frac{\Delta_{\rm sc}^{0}(x)}{2}(\cos k_x a -\cos k_y a),  &
\theta_{yrz}<\theta<\frac{\pi}{2}-\theta_{yrz}\\
\\
\displaystyle
0, &
{\text{\rm otherwise.}}
\end{array}\right.
\label{eq:sccut}
\end{equation}
With this $\Delta_{\rm sc}$ cutoff, the gaps in the nodal direction are largely unaffected, and the temperature dependence of the corresponding B$_{2g}$ peaks follows exactly as in Fig.~\ref{fig:temperature1} (lower frame).
This is not the case for the B$_{1g}$ antinodal results, where the superconductivity has been removed.  These results are shown in Fig.~\ref{fig:temperature2}.  The open blue triangles are for the $x$=0.16 case from Fig.~\ref{fig:temperature1}, which had no $\Delta_{\rm sc}$ cutoff.  Upon inclusion of the cutoff, the results change drastically to the solid blue circles, which show no temperature dependence (due to the lack of temperature dependence assumed for the pseudogap).  Although not shown, all dopings below $x$=0.16 show the exact same result that the dominant peak is purely pseudogap and independent of $T$.  However, as the doping is increased above $x$=0.16, there is reconstruction of the Fermi pocket in the antinodal wings, shown in the inset of Fig.~\ref{fig:temperature2} for the $x$=0.18 case.  For these cases, the superconducting gap is present in the antinodal direction, and as a result, the temperature dependence of the B$_{1g}$ peaks returns.  This change in Fermi surface in the antinodal direction occurs over a relatively small doping change, $\Delta x$=0.01--0.02, and results in drastic modification to the temperature dependence of the B$_{1g}$ Raman spectra.  This gap cutoff could explain the recent data of Guyard \emph{et al.}, \cite{guyard:2008} which sees a large change in the B$_{1g}$ temperature dependence over a small range of doping.  
Furthermore, dopings which present little or no temperature dependence in the B$_{1g}$ spectra, despite a strong B$_{2g}$ nodal gap $\Delta_{N}$, are evidence of a pseudogap suppressing the antinodal part of the Fermi surface.  Experimental observation of this lack of temperature dependence continuing below $T_c$ is evidence to support the idea that the pseudogap line continues through to $T$=0 in the phase diagram, and does not terminate at the top of the dome.

\begin{figure}
 \centering
\includegraphics[width=0.96\linewidth]{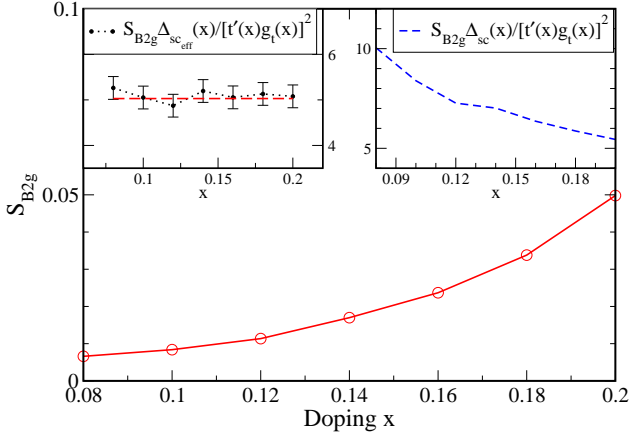}
\caption{ (Color online) Low frequency slope of the B$_{2g}$ Raman response,  $S_{B2g}=\left[\frac{\partial \chi^{\prime\prime}_{B2g}}{\partial \omega}\right]_{\omega=0}$, versus doping.  In the top right inset, doping dependent factors $[t^\prime(x)g_t(x)]^2$ have been removed to attempt to verify the low frequency scaling of $\chi^{\prime\prime}_{B2g} \propto \omega/\Delta_{\rm sc}$.  The top left inset shows that the scaling relation is changed to $\chi^{\prime\prime}_{B2g} \propto \omega/\Delta_{\rm sc_{eff}}$ in order to compare slopes across all dopings.  To guide the eye in the top left inset, we have included a horizontal line at the average value of the points and included error bars of $\pm$5$\%$ of the point values. }\label{fig:B2gslopes}
\end{figure}

Comparing Raman spectra across the phase diagram is complicated due to doping dependent changes to the vertex weightings, as well as modifications to the Fermi surface.  To further emphasize this point, we calculate the low frequency slope of the nodal Raman response, $S_{B2g}=\left[\frac{\partial \chi_{B2g}^{\prime\prime}}{\partial \omega}\right]_{\omega=0}$, across the doping phase diagram.  In BCS theory of $d$-wave superconductors, this low frequency dependence goes linear with frequency as $\chi_{B2g}^{\prime\prime} \propto \omega/\Delta_{\rm sc}$.  Since for most dopings there is little impact due to the pseudogap on the nodal direction response, the effects of the vertex component, as well as Fermi pocket size, can be seen clearly.  Fig.~\ref{fig:B2gslopes} shows these slopes versus doping as open red circles with the solid red line a guide to the eye.  Taken directly, there is a factor of 5 difference across the region shown.  We can largely remove the doping dependent vertex factors by scaling out $[t^\prime(x)g_t(x)]^2$, and can compensate for the superconducting gap by multiplying by $\Delta_{\rm sc}(x)$.  This results in the dashed blue curve in the upper right inset of Fig.~\ref{fig:B2gslopes}.  With the factors removed there remains a doping dependence, which varies by a factor of 2 across the doping range shown.  To compensate for the effect of the pocket size, we replace the $\Delta_{\rm sc}(x)$ factor with a $\Delta_{\rm sc_{eff}}(x)$ factor, which is shown in the top left inset of Fig.~\ref{fig:B2gslopes}.  To within 5\% in the values here, the simple BCS low frequency scaling is maintained since this is purely nodal physics and should be relatively unaffected by the higher energy pseudogap scale.  This does illustrate however, that the indirect effect of the pseudogap of modifying the Fermi surface has an important and strong impact on the low frequency B$_{2g}$ spectra.  This shows that it is the value of the superconducting gap on the Fermi pocket that is important, and not its value at other points in the Brillouin zone. 

\begin{figure}
 \centering
\includegraphics[width=0.96\linewidth]{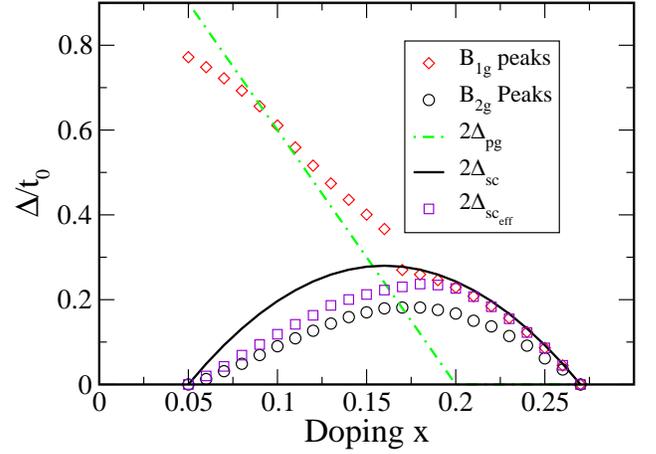}
\caption{(Color online) Gaps versus doping.  Shown are input values of $2\Delta_{\rm sc}(x)$ (solid black) and $2\Delta_{\rm pg}(x)$ (dash-dotted green) given by Eqs.~(\ref{eqn:scdome}) and (\ref{eqn:pgline}), respectively.  Open red diamonds are the B$_{1g}$ peaks extracted at zero temperature.  Open black circles are the B$_{2g}$ peaks at zero temperature.  The open purple squares give the effective superconducting gap, $2\Delta_{\rm sc_{eff}}(x)$, as described in Fig.~\ref{fig:sceff}.  In the overdoped region, the B$_{1g}$ peaks fall on the $\Delta_{\rm sc_{eff}}(x)$ line, while the B$_{2g}$ peaks fall well below.}\label{fig:phased}
\end{figure}

Finally, we wish to examine the progression of zero temperature peaks in the Raman response as the doping is varied over the phase diagram.  Fig.~\ref{fig:phased} shows such a plot, which includes the input $2\Delta_{\rm sc}(x)$ dome (solid black curve) and $2\Delta_{\rm pg}(x)$ line (dash-dotted green line).  Also included is $2\Delta_{\rm sc_{eff}}(x)$ (open purple squares), the maximum gap value on the Fermi pocket, from Fig.~\ref{fig:sceff}.  All zero temperature calculations are for a superconducting gap which exists over the entire Fermi surface and not just on the pocket, as in Eq.~(\ref{eqn:scgap}).  The open red diamonds mark B$_{1g}$ peaks, while open black circles mark B$_{2g}$ peaks.
One first notes that the B$_{2g}$ peaks lie substantially below the input $2\Delta_{\rm sc}$ dome and follow a curved shape much like the $2\Delta_{\rm sc_{eff}}(x)$ modified dome.  In practice, the $2\Delta_{\rm sc_{eff}}$ point should always lie beyond the B$_{2g}$ peak at a point of inflection (as mentioned in discussion of Fig.~\ref{fig:raman1}).  Here, if one were to divide $2\Delta_{\rm sc_{eff}}(x)$ by the B$_{2g}$ peak values, you would find a roughly constant value of $\approx 1.4$.  The B$_{1g}$ peaks show additional features.  For dopings above $x_c$=0.20, the B$_{1g}$ peaks fall exactly on the $2\Delta_{\rm sc_{eff}}$ dome.  Just below $x_c$, for dopings $x$=0.17$\rightarrow$0.19, the B$_{1g}$ peaks lie just above the $2\Delta_{\rm sc_{eff}}$ dome, due to the presence of a small pseudogap.  For these dopings, there is still Fermi surface in the antinodal region in the form of small `loops' which are nearest approach surfaces for the $E_s^+$ band, as is illustrated in the inset of Fig.~\ref{fig:temperature2} for the $x$=0.18 case.  There is also a jump in this curve between $x$=0.16 and $x$=0.17.  For dopings above $x$=0.17, the B$_{1g}$ spectral peaks coincide with the $E_s^+$ energy on the nearest approach surface in the antinodal direction.  For dopings below $x$=0.17, the $E_s^+$ `loops' disappear altogether.  This results in a shift from the $E_s^+$ band to the $E_s^-$ band, which happens to have a higher energy of nearest approach in the antinodal direction.  As a result, the B$_{1g}$ peaks jump up in energy as $x$ is changed from 0.17 to 0.16.  We will refer to this jump at $x$=0.16 as being at the onset doping, $x_{onset}$, the doping below which the system is dominated by pseudogap-created Fermi pockets.

We now have the image of three distinct regions: $x<x_{onset}$, which is dominated by the pseudogap and the Fermi pockets; $x_{onset}<x<x_c$,  which is a region wherein strong Fermi surface reconstruction occurs; and finally $x>x_c$, where the pseudogap is zero and the system is dominated by the $\Delta_{\rm sc}$ energy scale with a large tight-binding Fermi surface described by a Fermi liquid model. 
Of these three regions, the one that is most complicated to understand quantitatively is the middle range, where the topology of the Fermi surface becomes particularly complex and includes Fermi pockets as well as other pieces of Fermi surface with complex geometry around the antinodal points as seen in the inset of Fig.~\ref{fig:temperature2}.  It is however precisely this region where the B$_{1g}$ Raman scattering cross-section shows the most rapid variation, as seen in Fig.~\ref{fig:restruct}, and has the greatest promise for future experimental investigation.  A second interesting feature to note about the phase diagram traced in Fig.~\ref{fig:phased} is that the values of the position of the B$_{1g}$ Raman peaks (open red diamonds) fall below the input pseudogap values (green dash-dotted curve) in the highly underdoped region near the lower end of the superconducting dome.  These reduced values coincide with the value of the $E_s^-$ band on the antinodal nearest approach surface.  The physics behind this reduction in peak position relates to the shape of the B$_{1g}$ Raman vertex and its overlap with the Fermi pocket as is seen in Fig.~\ref{fig:ud}.  The Fermi pockets have shrunken so much that the peak in the B$_{1g}$ Raman cross-section comes from sampling a momentum region which is displaced from the antinodal direction towards the nodal direction and in this region, the pseudogap is reduced from its full amplitude $\Delta_{\rm pg}^0(x)$, by the modulating factor $[\cos (k_x a) - \cos (k_y a)]$ of Eq.~(\ref{eqn:pggap}).

\begin{figure}
 \centering
\includegraphics[width=0.96\linewidth]{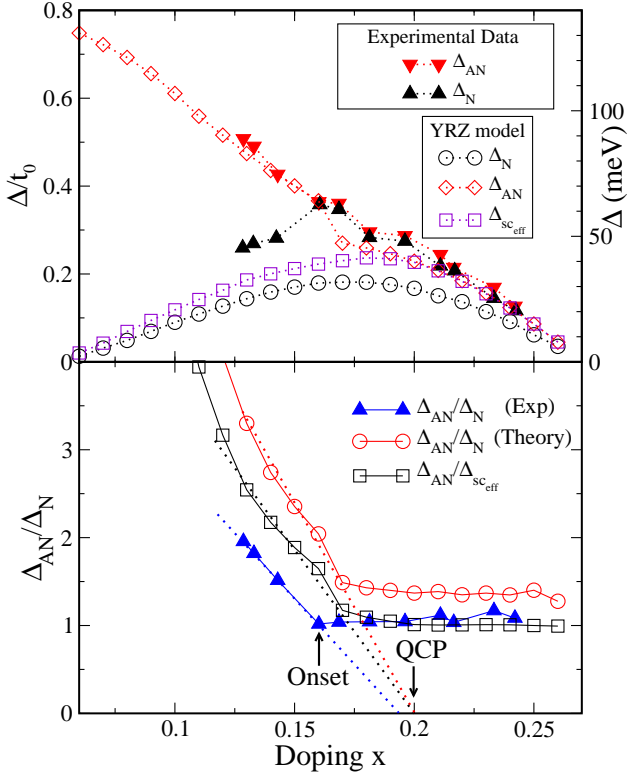}
\caption{(Color online) Top frame: The data of Guyard \emph{et al.} [Ref.~\cite{guyard:2008} Fig.~2(a)] for $\Delta_{AN}$ and $\Delta_N$ is shown in solid points (right hand axis) while the YRZ results for $\Delta_{AN}$, $\Delta_N$ and $\Delta_{\rm sc_{eff}}$ are shown in open points (left hand axis).  Bottom frame: Comparison of the antinodal to nodal gap ratio $\Delta_{AN}/\Delta_N$ for experiment (solid blue triangles) and YRZ model (open red circles) as well as to the YRZ result for $\Delta_{AN}/\Delta_{\rm sc_{eff}}$ (open black squares).}\label{fig:ratio}
\end{figure}

Fig.~\ref{fig:ratio} contains a comparison of the results of our calculations to the experimental data of Guyard \emph{et al.}, taken from Ref.~\cite{guyard:2008}.  The top frame contains two y-axis scales.  The left scale is for our calculation, with dimensionless quantity $\Delta/t_0$ and applies to the open points, while the right hand scale has units of meV and applies to the experimental data marked with filled points.
The experimental and theoretical scales are shown with equivalent values for the assumed $t_0=175$ meV in the YRZ model.  It is clear on first glance that the energy scale for our calculations agrees well with experiment despite not being rigorously fit to this specific experimental system of Hg-1201.
The experimental data sets show that $\Delta_{N}$ (solid black upward pointing triangles) and $\Delta_{AN}$ (solid red downward pointing triangles) have very similar values above some doping around $x\approx0.16$.  This feature is not captured in our calculation for the $\Delta_{N}$ (open black circles) and $\Delta_{AN}$ (open red diamonds), as $\Delta_{N}$ maintains a lower value than $\Delta_{AN}$ for all dopings.  We note however that $\Delta_{sc_{eff}}(x)$ (open purple squares) has similar values to $\Delta_{AN}$ (open red diamonds) in the overdoped region.  This illustrates that the actual $\Delta_{N}$ doping dependence, as seen in experiment, may not necessarily follow a straightforward superconducting dome, but rather an effective dome, given by $\Delta_{\rm sc_{eff}}(x)$, which includes Fermi surface restructuring effects due to the pseudogap.
Our parameters could be improved by scaling down our energies, which could equivalently correspond to $t_0$ being smaller by a factor of $\approx 1.1$. 

In the lower frame of Fig.~\ref{fig:ratio}, we plot the experimental ratio $\Delta_{AN}/\Delta_{N}$ as solid blue triangles.  We can clearly see two doping dependent regions of behaviour in this data: $x<x_{onset}$ and $x>x_{onset}$, where $x_{onset}$ marks the change in slope around $x$=0.16.  We seek to distinguish the difference between $x_{onset}$ and $x_c$ in this data.  
The antinodal to nodal gap ratio, $\Delta_{AN}/\Delta_{N}$  (open red circles) as well as $\Delta_{AN}/\Delta_{\rm sc_{eff}}$ (shown in  open black squares) are shown for the YRZ model.  The $\Delta_{AN}/\Delta_{N}$ calculations also show two distinct regions: $x<x_{onset}$ and $x>x_{onset}$, where $x_{onset}$=0.17.  For dopings above $x_{onset}$, the $\Delta_{AN}$ and $\Delta_{N}$ curves do not have the same value, and instead, have a roughly constant ratio of $\approx 1.4$.  When we instead normalize $\Delta_{AN}$ by $\Delta_{\rm sc_{eff}}$, resulting in a constant ratio of $1$ for $x>0.16$, we find better agreement with the experimental observation of the $\Delta_{AN}/\Delta_{N}$ ratio.

An important point is that all three data sets (2 theoretical, 1 experimental) in the lower frame of Fig.~\ref{fig:ratio} show only two distinct regions of doping dependent behaviour: $x<x_{onset}$ and $x>x_{onset}$.  There is virtually no signature of $x_c$ in this data.  
We can analyze this further, since the $x_{onset}$ marks the dominance of $\Delta_{\rm pg}$ over $\Delta_{\rm sc_{eff}}$ in the phase diagram of Fig.~\ref{fig:phased}, all $x<x_{onset}$ are dominated by the pseudogap.  If we assume that the pseudogap energy scale maintains linearity, we can trace a simple linear fit of the points just below $x_{onset}$ and extend that line above $x_{onset}$.  This will result in an estimate of the location of the doping at which the pseudogap vanishes, and hence, the doping where one might look for evidence of a zero temperature QCP.  Indeed this extrapolation gives $x_c\approx 0.2$, as was input into the theory, and it provides an estimate from experiment that $x_c\approx 0.19$.  It is with this simple interpretation that we reestablish the existence of three regions of importance in this data: $x<x_{onset}$ (where $\Delta_{\rm pg}>\Delta_{\rm sc_{eff}}$), $x_{onset}<x<x_c$ (where $\Delta_{\rm pg}<\Delta_{\rm sc_{eff}}$) and $x>x_c$ (where $\Delta_{\rm pg}=0$).

\section{conclusions}
The YRZ model provides a formalism whereby a pseudogap opening up about the AFBZ boundary reconstructs a large Fermi surface into small Fermi pockets as the doping progresses towards the Mott insulating state.  Comparison of the YRZ results to recent Raman spectra on Hg-1201 gives excellent qualitative agreement, and establishes that its quantum critical point, associated with pseudogap formation, falls at doping $x$=0.19 inside the superconducting dome and is consistent with a zero temperature transition.  We have shown that if the superconducting gap is only present in the Fermi pocket region (nodal direction) then the B$_{1g}$ peaks show no temperature dependence.  However, beyond some critical doping, the Fermi surface reconstructs in the antinodal region of the Brillouin zone resulting in a large change in the B$_{1g}$ peak temperature dependence over a small change in doping.  We also understand the loss of B$_{1g}$ spectral amplitude in the superconducting state to be an effect of the presence of a dominant normal state background in the antinodal direction which grows in strength with underdoping.

Tracing the major peak in each of the B$_{1g}$ and B$_{2g}$ response curves as a function of doping creates an apparent phase diagram in the YRZ model which illustrates the indirect impact of the presence of the Fermi pocket.  The effect is that the Raman spectra can only sample the largest value of the gap on the Fermi pocket which is at the edge, towards the antinodal direction.  This results in an effective superconducting dome $\Delta_{\rm sc_{eff}}(x)$, which can be significantly smaller than the input gap.  This occurs for dopings where the pseudogap is strong, resulting in Fermi pockets which become smaller on approach to the antiferromagnetic state.  We have also analyzed the low frequency slope of the B$_{2g}$ polarization, taking into account the Fermi surface restructuring (pocket size) effect by replacing $\Delta_{\rm sc}$ with the effective gap, $\Delta_{\rm sc_{eff}}$.

Through this work we have identified three regions of doping dependence in the phase diagram:  the underdoped pseudogap-dominated region, the overdoped superconductivity-dominated Feremi liquid region, and the intermediate mixed region wherein the presence of pseudogap begins to erode the Fermi surface.  In the case of Raman spectra, this intermediate region may be seen best in the temperature dependence of the antinodal peaks, but also as a sudden jump of the antinodal peak frequency with doping.

Although our results are largely concerned with the formation of Fermi pockets, we understand that there are similar models of the pseudogap state which result in the formation of Fermi arc segments rather than Fermi pockets.  The qualitative results given here will remain unchanged, as they require: 1) some disappearance of Fermi surface in the antinodal direction due to the presence of a pseudogap, 2) a dominant superconducting gap value given by the maximum value on the Fermi surface.  These two requirements should still be present in models which result in arcs rather than pockets.
\begin{acknowledgments}
  This work has been supported by the Natural Sciences and
Engineering Council of Canada (NSERC) and the Canadian Institute
for Advanced Research (CIFAR).
\end{acknowledgments}
\bibliographystyle{apsrev}
\bibliography{bib}

\end{document}